\documentclass[12pt]{article}
\usepackage{graphics}
\begin{document}
\markright{Unexpectedly large surface gravities for acoustic
horizons?\hfil}
\def\Universita{Universit\`a}
\title{\bf \LARGE
Unexpectedly large surface gravities for acoustic horizons?}
\author{
{\Large Stefano Liberati}
\\[2mm]
{\small \it International School for Advanced Studies
(ISAS/SISSA),}
\\ {\small \it Via Beirut 2--4, 34014 Trieste, Italy}
\\ {\small and}
\\ {\small \it Instituto Nazionale di Fisica Nucleare (INFN),}
\\ {\small \it sezione di Trieste}
\\ and
\\ {\Large Sebastiano Sonego}
\\[2mm]
{\small \it {\Universita} di Udine, Via delle Scienze 208,
33100 Udine, Italy}
\\ and
\\ {\Large Matt Visser}
\\[2mm] {\small \it Physics Department, Washington University,}
\\ {\small \it Saint Louis, Missouri 63130-4899, USA}}
\date{{\small 28 March 2000; Revised 30 May 2000; \LaTeX-ed \today }}
\maketitle
\vfill
\hrule
\bigskip
\centerline{\underline{E-mail:}
{\sf liberati@sissa.it}}
\centerline{\underline{E-mail:}
{\sf sebastiano.sonego@uniud.it}}
\centerline{\underline{E-mail:}
{\sf visser@kiwi.wustl.edu}}
\bigskip
\centerline{\underline{Homepage:}
{\sf http://www.sissa.it/\~{}liberati}}
\centerline{\underline{Homepage:}
{\sf http://www.physics.wustl.edu/\~{}visser}}
\bigskip
\centerline{\underline{Archive:}
{\sf gr-qc/0003105}}
\bigskip
\hrule
\clearpage
\begin{abstract}

Acoustic black holes are fluid dynamic analogs of general
relativistic black holes, wherein the behaviour of sound
waves in a moving fluid acts as an analog for scalar fields
propagating in a gravitational background. Acoustic
horizons, which are intimately related to regions where the
speed of the fluid flow exceeds the local speed of sound,
possess many of the properties more normally associated with
the event horizons of general relativity, up to and
including Hawking radiation. Acoustic black holes have
received much attention because it would seem to be much
easier to experimentally create an acoustic horizon than to
create an event horizon. In this note we wish to point out
some potential difficulties (and opportunities) in actually
setting up an experiment that possesses an acoustic horizon.
We show that in zero-viscosity, stationary fluid flow with
generic boundary conditions, the creation of an acoustic
horizon is accompanied by a formally infinite ``surface
gravity'', and a formally infinite Hawking flux. Only by
applying a suitable non-constant external body force, and
for very specific boundary conditions on the flow, can these
quantities be kept finite. This problem is ameliorated in
more realistic models of the fluid. For instance, adding
viscosity always makes the Hawking flux finite (and
typically large), but doing so greatly complicates the
behaviour of the acoustic radiation --- viscosity is
tantamount to explicitly breaking ``acoustic Lorentz
invariance''. Thus, this issue represents both a difficulty
and an opportunity --- acoustic horizons may be somewhat
more difficult to form than naively envisaged, but if
formed, they may be much easier to detect than one would at
first suppose.
\end{abstract}
\vfill
\def\Box{\nabla^2}
\def\d{{\mathrm d}}
\def\x{{\vec x\,}}
\def\p{{\vec p\,}}
\def\k{{\vec k\,}}
\def\v{{\vec v\,}}
\def\n{{\vec n\,}}
\def\half{{1\over2}}
\def\quarter{{1\over4}}
\def\L{{\cal L}}
\def\sech{\hbox{sech}}
\def\SIZE{1.00}
\def\vecV{{\vec{f}_{\mathrm{viscous}}}}
\def\ie{{\em i.e.\/}}
\def\eg{{\em e.g.\/}}
\def\etc{{\em etc.\/}}
\def\etal{{\em et al.\/}}
\def\Nordstrom{Nordstr\"om}
\def\Painleve{Painlev\'e}
\clearpage
\section{Introduction}
\setcounter{equation}{0}
Acoustic black holes are very useful toy models that share
many of the fundamental properties of the black holes of
general relativity, while having a very clear and clean
physical interpretation in terms of ordinary
non-relativistic fluid mechanics
\cite{Unruh81,Jacobson91,Jacobson93,Visser93,Unruh94,Visser98,Visser97,Visser99}.
The fundamental idea is that sound waves propagating in a
flowing fluid share many of the formal properties of
massless scalar fields propagating in a general-relativistic
curved spacetime. Indeed, the propagation of acoustic
disturbances in a flowing fluid is described by a spacetime
metric with Lorentzian signature, the ``acoustic metric'',
which is built up algebraically out of the density,
velocity, and local speed of sound of the fluid. When the
flow is such that there is a surface where the normal
component of the fluid velocity equals the speed of sound,
the acoustic metric possesses the properties that
characterize a black hole spacetime in general relativity,
and such a surface is therefore called ``acoustic horizon''.

As emphasized in \cite{Visser98,Visser97}, acoustic black
holes share all the {\em kinematic\/} aspects of
relativistic black holes, but do not share in the {\em
dynamic\/} aspects. In particular, acoustic black holes
exhibit Hawking radiation from the acoustic horizon, giving
rise to a quasi-thermal bath of phonons with temperature
proportional to the ``surface gravity'' (related to the {\em
physical\/} acceleration of the fluid as it crosses the
acoustic horizon), but they exhibit no simple analog of the
Bekenstein--Hawking entropy (since that is a dynamical
effect intimately related to the existence of the Einstein
equations in general relativity).

One of the reasons why acoustic black holes are so popular
is that it seems that the prospects for experimentally
building an acoustic horizon are much better than for a
general relativistic event horizon. An early estimate can be
found in \cite{Unruh81}, and related comments are to be
found in \cite{Visser98,Visser97}. Additionally, an
impressive body of work is due to Volovik and collaborators,
who have extensively studied the experimental prospects for building such
a system using superfluids such as ${}^3$He and ${}^4$He
\cite{Volovik:1997xi,Kopnin:1998jy,Volovik:1998de,Volovik:1997pf,%
Jacobson:1998ms,Jacobson:1998he,Volovik:1999zs,Volovik:1999fc,%
Volovik:1999cn,Volovik:2000}. These particular implementations of
acoustic geometry make extensive use of the two-fluid model of
superfluidity, whereas in this paper we will be focussing on a
conceptually simpler one-fluid model; accordingly, some important
technical details will differ.  For yet another physical
implementation of acoustic geometries, Garay {\etal} have investigated
the technical requirements for implementing an acoustic horizon in
Bose--Einstein condensates~\cite{Garay}, and some of the perils and
pitfalls accompanying acoustic black holes have been discussed in
Jacobson's mini-survey~\cite{Mini-survey}.

Another attractive feature of acoustic black holes is that
they seem to be generic, and that they illustrate an
important aspect of Lorentz invariance. For instance, it is
now known, due to the work of Nielsen and
collaborators~\cite{Nielsen78,Nielsen83a,Nielsen83b}, that
in renormalizable non-Lorentz-invariant quantum field
theories, Lorentz invariance is often an infrared fixed
point of the renormalization group equations. Thus, Lorentz
invariance can emerge as a symmetry in the low-energy limit
even if the underlying physics is not explicitly
Lorentz invariant. Similarly, in acoustic black holes the
underlying physics is explicitly classical and Newtonian,
but the physics of sound propagation nevertheless exhibits a
low-frequency approximate Lorentzian
symmetry~\cite{Visser98,Visser97}.

In this note we wish to point out a potential difficulty and
an opportunity --- we shall demonstrate that there is a
regularity issue that becomes serious at the acoustic
horizon. Either the Hawking temperature is formally infinite
(which is the generic situation), or there must be a very
precise relationship between an {\em external\/} body force
that must be applied to the fluid as it crosses the acoustic
horizon and the extrinsic geometry of the latter. If this
condition is not satisfied the ``surface gravity'' formally
diverges, as well as the corresponding Hawking temperature.
Similarly, the acceleration and density gradient of the
fluid at the horizon are formally infinite. For a specified
external force, such divergences are generic, in the sense
that they are present for almost all flows, except --- in
some cases --- for a set of measure zero that satisfy very
special boundary conditions. However, in the case of a
constant force (including zero force), which is perhaps the
most interesting one from the point of view of laboratory
simulations, no boundary conditions exist that correspond to
an everywhere regular flow.

On the one hand this result suggests that detecting the
acoustic Hawking effect should be very easy; on the other
hand it implies that the naive analysis (which demands that
both the vorticity and the viscosity be zero) should in some
way be modified near the acoustic horizon, at least when the
external forces are such that a formal divergence will
certainly occur. For instance, adding finite viscosity to
the fluid equations is sufficient in order to regulate the
surface gravity and Hawking temperature for any choice of
external force --- though finite they can remain large, and
can be much larger than naively expected.

\section{Basic equations and assumptions}
\setcounter{equation}{0}

The acoustic model of Lorentzian geometry arises from the
description of the deceptively simple phenomenon of the
propagation of sound waves in a flowing fluid. Let us
therefore recall the fundamental equations of fluid
dynamics, \ie, the equation of continuity
\begin{equation}
{\partial\rho\over\partial t}
+ \vec\nabla\cdot\left(\rho\, \v \right)= 0,
\label{E:continuity}
\end{equation}
and the Euler equation
\begin{equation}
\rho\;\vec{a}=\vec{f},
\label{E:euler0}
\end{equation}
where
\begin{equation}
\vec{a}={\partial\v\over\partial t} +
(\v \cdot \vec\nabla) \v
\label{E:acc}
\end{equation}
is the fluid acceleration, and $\vec f$ stands for the force
density --- the sum of all forces acting on the fluid per
unit volume. We shall assume that the external forces
present are all gradient-derived (possibly time-dependent)
body forces, which for simplicity we lump together in a
generic term $-\rho\,\vec\nabla\Phi$. In addition to the
external forces, $\vec f$ contains a contribution from the
pressure of the fluid and, possibly, a term coming from
viscosity. Thus, equation (\ref{E:euler0}) takes the
Navier-Stokes form
\begin{equation}
\rho\left({\partial\v\over\partial t} +
(\v \cdot \vec\nabla) \v\right)
=- \vec\nabla p - \rho\;\vec\nabla\Phi +
{\vecV},
\label{E:euler}
\end{equation}
where
\begin{equation}
{\vecV}=\eta\nabla^2 \v+
\left(\zeta+{1\over 3}\,\eta\right)
\vec{\nabla}(\vec{\nabla}\cdot\v)
\end{equation}
represents the force due to viscous processes, the coefficients $\eta$
and $\zeta$ giving the dynamic and bulk viscosity, respectively
\cite{ll,kundu}.

In deriving the acoustic geometry, one usually makes a number of
technical assumptions.
\begin{itemize}
\item
The first assumption is that the fluid has a barotropic
equation of state, that is, the density $\rho$ is a function
only of the pressure $p$, so
\begin{equation}
\rho = \rho(p).
\end{equation}
This guarantees that (\ref{E:continuity}) and (\ref{E:euler}) are a
closed set of equations. We shall consequentely define the speed of
sound as
\begin{equation}
c^2 = {\d p\over\d\rho}.
\end{equation}
\item
The second assumption is that we have a vorticity-free flow, {\em
i.e.\/}, that $\vec\nabla\times\v=\vec 0$. This condition is generally
fulfilled by the superfluid components of physical superfluids.
\item
A third assumption, often made in the existing literature on acoustic
geometries, is a viscosity-free flow. Although this is quite a
realistic condition for superfluids we shall see that the presence or
absence of viscosity can mark a sharp difference in the behaviour of
the phonon radiation from acoustic horizons.
\end{itemize}

These assumptions are sufficient conditions under which an
acoustic metric can be written. However, since the following
analysis is independent of the introduction of the acoustic
geometry (although motivated by it, of course), we shall try
to be as general as possible, making use of them only
progressively, as they are needed in order to have an
analytically tractable system.

\section{Regularity conditions at ergo-surfaces}
\setcounter{equation}{0}
\label{S:ergo-surface}
Let us start by establishing a useful mathematical identity.
If we write $\v=v\n$, where $\n$ is a unit vector and $v\geq
0$, then
\begin{equation}
\vec\nabla\cdot\v={\d v\over\d n} + v \; K,
\label{E:useful}
\end{equation}
where $\d /\d n=\n\cdot\vec\nabla$ and
$K=\vec\nabla\cdot\n$. If the Frobenius condition is
satisfied,\footnote{
The Frobenius condition is $\v\cdot\vec\nabla\times\v=0$, or
equivalently $\n\cdot\vec\nabla \times\n=0$. This is
sometimes phrased as the statement that the flow has zero
``helicity''. The Frobenius condition is satisfied whenever
there exist a pair of scalar potentials such that $\v =
\alpha \, \vec \nabla \beta$, in which case the velocity
field is orthogonal to the surfaces of constant $\beta$. In
view of this fact the velocity field is said to be a
``surface-orthogonal vector field''.}
then there exist surfaces orthogonal to the fluid flow. In
this situation, $K$ admits a geometrical interpretation as
the trace of the extrinsic curvature of these surfaces. It
must be noted that, although zero vorticity is a sufficient
condition for this to happen, it is not necessary.

We now focus our attention on the component of the fluid
acceleration along the flow, $a_n=\vec{a}\cdot\n$. This can
be obtained straightforwardly by projecting the
Navier-Stokes equation (\ref{E:euler}) along $\n$:
\begin{equation}
\rho\, a_n
=
- c^2\,{\d\rho\over\d n}
-\rho\,{\d\Phi\over\d n}
+ \n\cdot{\vecV},
\label{E:aaa}
\end{equation}
where we have used the barotropic condition.

Next, we rewrite the continuity equation as
\begin{equation}
{\partial\rho\over\partial t}
+v\,{\d\rho\over\d n}
+\rho\left({\d v\over\d n}+vK\right)=0,
\label{E:urca}
\end{equation}
where the identity (\ref{E:useful}) has been used. We can
express $\d v/\d n$ in terms of $a_n$ noticing that, by the
definition (\ref{E:acc}) of $\vec{a}$,
\begin{equation}
a_n={\partial v\over\partial t}+v\,{\d v\over\d n}.
\label{E:boh}
\end{equation}
Thus, equation (\ref{E:urca}) can be rewritten as
\begin{equation}
\rho \; a_n
=
- v^2\,{\d\rho\over\d n}
-\rho v^2K
+ \rho\,{\partial v\over\partial t}
- v\,{\partial \rho\over\partial t}\;.
\label{E:bbb}
\end{equation}
Equations (\ref{E:aaa}) and (\ref{E:bbb}) can be solved for
both $a_n$ and $\d\rho/\d n$, obtaining:
\begin{equation}
a_n
=
{1\over c^2-v^2}
\left[
  v^2\left({\d\Phi\over\d n}- c^2K-
  {1\over\rho}\,\n\cdot{\vecV}\right)
+ c^2\left({\partial v\over\partial t}
           -{v\over\rho}\,{\partial\rho\over\partial t}
     \right)
\right];
\label{E:an}
\end{equation}
\begin{equation}
{\d\rho\over\d n}
=
{1\over c^2-v^2}
\left[
-\rho\left({\d\Phi\over\d n}
-v^2K
- {1\over\rho}\,\n\cdot{\vecV}\right)
-\rho\,{\partial v\over\partial t}
+v\,{\partial\rho\over\partial t}
\right].
\label{E:gradrho}
\end{equation}
In general we see that there is risk of a divergence in the
acceleration and the density gradient as $v \to c$, which
indicates that the ergo-surfaces\footnote{
In general relativity the $v \to c$ surface would be called
an ``ergosphere'', however proving that this surface
generically has the topology of a sphere is a result special
to general relativity which depends critically on the
imposition of the Einstein equations. In the present fluid
dynamics context there is no particular reason to believe
that the $v\to c$ surface would generically have the
topology of a sphere and we prefer the more non-committal
term ``ergo-surface''.}
(the boundaries of ergo-regions) must be treated with some
delicacy. The fact that gradients diverge in this limit is
the key observation of this paper; we shall demonstrate that
this has numerous repercussions throughout the physics of
acoustic black holes.

Since $v^2=c^2$ at the ergo-surface it is evident that the
acceleration and the density gradient both diverge, unless
the condition
\begin{equation}
{\d\Phi \over \d n}
- c^2 K
-  {1\over\rho}\,\n\cdot{\vecV}
+ {\partial v\over\partial t}
- {c\over\rho}\,{\partial\rho\over\partial t}=0
\label{E:main}
\end{equation}
is satisfied on the ergo-surface. Equation (\ref{E:main}) is
therefore a relationship that must be satisfied in order to
have a physically acceptable model. Of course, it is only a
{\em necessary\/} condition, because $a_n$ and $\d\rho/\d n$
may diverge at the ergo-surface even when (\ref{E:main}) is
fulfilled, if the quantities in square brackets in the right
hand sides of (\ref{E:an}) and (\ref{E:gradrho}) tend to
zero slower than $c^2-v^2$ as one approaches the
ergo-surface.

For a stationary, non-viscous flow, (\ref{E:main}) reduces
to
\begin{equation}
{\d\Phi \over \d n}=c^2 K,
\label{E:fine-tuning}
\end{equation}
where again $d\Phi/\d n$, $c$, and $K$ are evaluated at a
generic point on the ergo-surface. Thus, in this case it
seems that a special fine-tuning of the external forces is
needed in order to keep the acceleration and density
gradient finite at the ergo-surface. If the condition
(\ref{E:fine-tuning}) is not fulfilled but still
${\vecV}=\vec{0}$, the flow cannot be stationary. Near the
ergo-surface, an instability will make the time derivatives
in (\ref{E:main}) different from zero, so that they could
compensate the mismatch between the two sides of
(\ref{E:fine-tuning}). More realistically, we shall see
later that for a given potential, either no horizon forms,
or the flow tries to assume a configuration in which
(\ref{E:fine-tuning}) is automatically satisfied.

\section{Regularity conditions at horizons}
\setcounter{equation}{0}
\label{S:horizon}
If we now look at the ``surface gravity'' of an acoustic
black hole it is most convenient to first restrict attention
to a stationary flow. Defining a notion of surface gravity
for non-stationary flows is easier in fluid mechanics than
in general relativity, but is still sufficiently messy to
encourage us to make this simplifying
assumption~\cite{Visser98}. For additional technical
simplicity we shall further assume that at the acoustic
horizon (the boundary of the trapped region) the fluid flow
is normal to the horizon. Under these circumstances the
technical distinction between an ergo-surface and an
acoustic horizon vanishes and we can simply define an
acoustic horizon by the condition $v=c$. (In complete
generality you would have to define an [apparent] acoustic
horizon as a surface for which the inward {\em normal
component} of the fluid velocity is everywhere equal to the
speed of sound~\cite{Visser98}; this adds extra layers of
technical complication to the discussion which in the
present context we have not found to be useful.) Then it can
be shown that the surface gravity\footnote{
Hereafter, we label all quantities evaluated at the horizon
with the index $H$.}
$g_H$ has two terms~\cite{Visser98,Visser97}, one coming
from acceleration of the fluid, the other coming from
variations in the local speed of sound. More precisely,
$g_H$ is given by the value attained by the quantity
\begin{equation}
g =
\half \left.{\d(c^2-v^2)\over\d n}\right.
=
\half \left.{\d c^2\over\d n}\right. - a_n
\end{equation}
at the acoustic horizon. (And note that $g$ is defined
throughout all space.)

Now we write, using equations (\ref{E:urca}) and
(\ref{E:boh}),
\begin{eqnarray}
{\d c^2\over\d n}
&=&
{\d c^2\over\d \rho} \; {\d\rho\over\d n}
\nonumber\\
&=&
-{\d^2 p\over\d \rho^2}
\left[
\rho\left(K + {1\over v}
{\d v\over\d n}\right)+ {1\over v} \,
{\partial\rho\over\partial t}
\right]
\nonumber\\
&=&
-{\d^2 p\over\d \rho^2}
\left[
\rho
\left(
K + {a_n\over v^2}-
{1\over v^2}\,{\partial\rho\over\partial t}
\right)
+{1\over v}\,{\partial\rho\over\partial t}
\right],
\end{eqnarray}
so we find, using (\ref{E:an}):
\begin{eqnarray}
g
&=&
{1\over c^2-v^2}
\left[
\left(c^2 + {\rho\over 2}\,{\d^2 p\over\d \rho^2} \right)
\left(
K v^2-{\partial v\over\partial t}+
{v\over\rho}\,{\partial\rho\over\partial t}
\right)
\right.
\nonumber\\
&&
\left.
-\left(v^2 + {\rho\over
2}\,{\d^2 p\over\d \rho^2}
\right)
\left({\d\Phi\over\d n} - {1\over\rho}\,\n\cdot{\vecV} \right)
\right].
\end{eqnarray}

Under the present assumptions, time derivatives vanish and
$v^2=c^2$ at the horizon, so it is now evident that the
surface gravity (as well as the acceleration and the density
gradient) diverges unless the condition
\begin{equation}
\left.{\d\Phi \over \d n}\right|_H
- c_H^2 K_H
-  {1\over\rho_H}\,\n_H\cdot({\vecV})_H
= 0
\label{E:main2}
\end{equation}
is satisfied. For a non-viscous flow (\ref{E:main2}) again
reduces to (\ref{E:fine-tuning}), and the same
considerations made about the acceleration and density
gradient apply.

Now all this discussion is predicated on the fact that
acoustic horizons actually form, and would be useless in the
case that some obstruction could be proven to prevent the
fluid from reaching the speed of sound. In order to deal
with this possibility we shall now check that at least in
some specific examples it is possible to form acoustic
horizons under the current hypotheses. For analyzing these
specific cases it is useful to first consider generic
stationary, spherically symmetric flow.

\section{Spherically symmetric stationary flow}
\setcounter{equation}{0}
\label{S:sss}

For simplicity, we now deal with the case of a spherically
symmetric stationary flow in $d$ space dimensions. Spherical
symmetry guarantees that the fluid flowlines are always
perpendicular to the acoustic horizon, and so we can ignore
the subtleties attendant on the distinction between horizons
and ergospheres~\cite{Visser98}. Additionally, for the time
being we shall assume the absence of viscosity,
${\vecV}=\vec{0}$.

For a spherically symmetric steady inflow, $\n$ is minus the
radial unit vector. Then $\d\n/\d n={\vec 0}$; also
\begin{equation}
{\d\over\d n} = -{\d\over\d r},
\end{equation}
and
\begin{equation}
K = - {d-1\over r}.
\end{equation}
{From} equation (\ref{E:acc}) it follows that $\vec{a}$ has
only the radial component, which coincides with $-a_n$ and
is
\begin{equation}
a = - v^2 \; { c^2 {(d-1)/r} - {\d\Phi/\d r}
\over
c^2 - v^2 }.
\label{E:accsph}
\end{equation}
This result could also be obtained directly, without the
general treatment of section \ref{S:ergo-surface}. For a
steady flow the continuity equation implies
\begin{equation}
\rho \; v \; r^{d-1} = J=\mbox{const}.
\label{E:continew}
\end{equation}
Taking the logarithmic derivative of the above equation one
gets
\begin{equation}
{\d \rho\over \d r} = - \rho {(d-1)\over r} - {\rho\over v}
\; {\d v\over\d r}.
\label{E:contsph}
\end{equation}
On the other hand the Euler equation (\ref{E:euler}) takes
in this case the form
\begin{equation}
\rho \; v \; {\d v\over\d r} = -c^2 \;
{\d \rho\over \d r} -
\rho {\d\Phi\over\d r},
\label{E:eulsph}
\end{equation}
where we have used the barotropic condition. Equations
(\ref{E:contsph}) and (\ref{E:eulsph}) can be combined to
give the useful result
\begin{equation}
v \; {\d v\over\d r} = c^2
\left( {(d-1)\over r} + {1\over v} \;
{\d v\over\d r} \right) - {\d\Phi\over\d r},
\label{E:glu}
\end{equation}
which allows one to easily compute the acceleration $a=v\,\d
v/\d r$ of the fluid for this specific case, recovering
equation\ (\ref{E:accsph}), and to obtain a differential
equation for the velocity profile $v(r)$:
\begin{equation}
{\d v\over\d r} =
- v \;
{
c^2 {(d-1)/r} -  (\d\Phi/\d r)
\over
c^2 - v^2
}.
\label{E:diffeq}
\end{equation}

When it comes to calculating $g$, the same analysis as
previously developed now yields
\begin{equation}
g=
{1\over c^2 - v^2}
\left[
\left(v^2 + {\rho\over 2}\,{\d^2 p\over\d \rho^2}\right)
{\d\Phi\over\d r}
-\left(c^2 + {\rho\over 2}\,{\d^2 p\over\d \rho^2}\right)
{v^2 (d-1)\over r}
\right].
\end{equation}
So the acceleration at the acoustic horizon, whose location
$r_H$ is the solution of the equation $v(r_H)^2=c(r_H)^2$,
formally goes to infinity unless the external body force
satisfies the condition
\begin{equation}
\left.{\d\Phi\over\d r}\right|_H - c^2_H \;
{(d-1)\over r_H} = 0.
\label{E:sss-fine-tuning}
\end{equation}

Any further analysis requires one to integrate the
differential equation (\ref{E:diffeq}). However, this can be
done only assigning an equation of state $p=p(\rho)$, and
integrating simultaneously equation (\ref{E:contsph}) in
order to get the dependence of $c$ on $r$. We consider such
a specific model in the next section.

\section{Constant speed of sound}
\setcounter{equation}{0}
\label{S:css}

In order to get further insight, let us consider the simple
case of a fluid with a constant speed of sound,
\begin{equation}
{\d^2 p \over \d\rho^2} = 0.
\label{E:c=const}
\end{equation}
It is easy to see that in this case, the condition
(\ref{E:sss-fine-tuning}) is also sufficient in order to
keep the physical quantities finite on the horizon. Consider
equation (\ref{E:diffeq}) and apply the Bernoulli--de
L'Hospital rule in order to evaluate $(\d v/\d r)_H$. One
gets
\begin{equation}
\left({\d v\over\d r}\right)^2_H=
-{1\over 2}\left(\left.{\d^2\Phi\over\d r^2}\right|_H
+{c^2(d-1)\over r_H^2}\right),
\label{E:dvdrh}
\end{equation}
so $(\d v/\d r)_H$ has a finite value. As a corollary of
(\ref{E:dvdrh}), we see that at the horizon one must have
\begin{equation}
\left.{\d^2\Phi\over\d r^2}\right|_H\leq
-{c^2(d-1)\over r_H^2};
\label{E:d2}
\end{equation}
so, in particular, no potential with a non-negative second
derivative can lead to a horizon on which $\d v/\d r$ is
finite.

\begin{figure}[htbp]
\vbox{
\hfil
\scalebox{0.25}{{\includegraphics{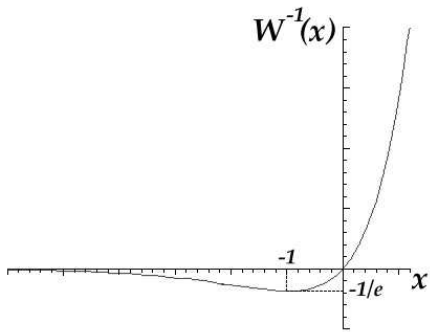}}}
\hfil
}
\bigskip
\caption{
The inverse of the Lambert function: $W^{-1}(x)=x\,{\rm
e}^x$.
}
\label{F:figlambert}
\end{figure}

With the assumption (\ref{E:c=const}), the differential
equation (\ref{E:diffeq}) for the velocity profile can be
easily integrated. Its general solution is
\begin{equation}
\frac{1}{2}
\left[
c^2 \;
\ln\left( \frac{v^2}{v^2_0}\right)-v^2+v^2_0
\right]
=
-c^2\;\left(d-1\right)\;\ln{\left(\frac{r}{r_0}\right)}
+ \Phi(r)-\Phi(r_0),
\label{E:d>1Kcos}
\end{equation}
where $r_0$ is arbitrary and $v_0$ is the speed of the fluid
at $r_0$.\footnote{Equation (\ref{E:d>1Kcos}) simply
expresses Bernoulli's theorem. Indeed, it can be written in
the form $v^2/2+\Phi(r)+h(v,r)=\mbox{const}$, where
$h=\int\d p/\rho$ can be found from (\ref{E:contsph}).} In
order to study the general properties of $v(r)$, it is
convenient to rewrite equation (\ref{E:d>1Kcos}) in the form
\begin{equation}
r^{2(d-1)}{\rm e}^{-2\Phi(r)/c^2}
W^{-1}\left(-v^2/c^2\right)
=r_0^{2(d-1)}{\rm e}^{-2\Phi(r_0)/c^2}
W^{-1}\left(-v_0^2/c^2\right),
\label{E:chisenefrega}
\end{equation}
where $W^{-1}$ is the inverse of the Lambert function
\cite{Lambert}, defined as $W^{-1}(x)=x \; {\rm e}^x$. Given
$r_0$ and $v_0$, equation (\ref{E:chisenefrega}) implies
that the solution $v(r)$ has two branches --- a subsonic and
a supersonic one. This follows immediately from the trivial
fact that, since $W^{-1}(-v_0^2/c^2)$ is negative, also
$W^{-1}(-v^2/c^2)$ is negative; then, from the plot in
figure \ref{F:figlambert} we see that there are two possible
values for $v^2$, one smaller and the other greater than
$c^2$.

We end this section with some remarks that are crucial for a
correct interpretation of the regularity condition
(\ref{E:sss-fine-tuning}). On rewriting (\ref{E:d>1Kcos}) or
(\ref{E:chisenefrega}) as
\begin{equation}
F(r,v;r_0,v_0)=0,
\label{E:F}
\end{equation}
we can represent the location $r_H$ of the horizon, for a
given potential $\Phi$ and given boundary data $(r_0,v_0)$,
as the solution of the equation
\begin{equation}
F(r_H,c;r_0,v_0)=0.
\label{E:rH}
\end{equation}
On the other hand, differentiating (\ref{E:F}) and comparing
with (\ref{E:diffeq}) we can rewrite the regularity
condition (\ref{E:sss-fine-tuning}) as
\begin{equation}
{\partial F\over\partial r}(r_H,c;r_0,v_0)=0.
\label{E:regcon}
\end{equation}
It is clear that, if we impose the boundary data
$(r_0,v_0)$, then (\ref{E:regcon}) expresses a fine-tuning
condition on $\Phi$ in order to have $\d v/\d r$ finite at
the horizon. However, we can reverse the argument and
consider the more realistic case in which one looks for a
physically acceptable flow compatible which an {\em
assigned\/} $\Phi$, {\em without\/} trying to force the
boundary condition $v(r_0)=v_0$ on the velocity profile. In
this case, equations (\ref{E:rH}) and (\ref{E:regcon}), when
solved simultaneously, give the location of the horizon,
$r_H$, {\em and\/} the value $v_0$ of the fluid speed at
$r_0$. Thus, requiring regularity of the flow for a given
potential amounts to solving an eigenvalue problem, while if
one insists on assigning a boundary condition for the speed,
a careful fine tuning of $\Phi$ is needed in order to avoid
infinite gradients. We stress, however, that although from a
strictly mathematical point of view both types of problems
can be considered, it is the first one that is relevant in
practice.

\section{Examples}
\setcounter{equation}{0}

We now consider some specific choices, both of $\Phi(r)$ and
of $v(r)$, in order to illustrate the general situation.

\subsection{Constant body force}

Let us begin with a constant body force, with the linear
potential
\begin{equation}
\Phi(r) = \kappa \; r,
\end{equation}
where $\kappa$ is a constant. Equation (\ref{E:d>1Kcos})
becomes, in this case,
\begin{equation}
\frac{1}{2}
\left[
c^2 \;
\ln\left( \frac{v^2}{v^2_0}\right)-v^2+v^2_0
\right]
=
-c^2\;\left(d-1\right)\;\ln{\left(\frac{r}{r_0}\right)}
+ \kappa(r-r_0),
\label{E:d>1Kcos2}
\end{equation}
and equation (\ref{E:chisenefrega}) can be rewritten
completely in terms of inverse Lambert functions:
\begin{eqnarray}
&&
W^{-1}\left(-{\kappa r\over c^2(d-1)}\right)^{2(d-1)}
W^{-1}\left(-{v^2\over c^2}\right)
\nonumber
\\
&&
\qquad
= W^{-1}\left(-{\kappa
r_0\over c^2(d-1)}\right)^{2(d-1)} W^{-1}\left(-{v_0^2\over
c^2}\right).
\label{E:diopoi}
\end{eqnarray}

Following the discussion at the end of section \ref{S:css},
we can regard (\ref{E:sss-fine-tuning}) as the equation for
the locations of $r_H$ where $\d v/\d r$ is finite. We have,
in this case,
\begin{equation}
r_H={c^2(d-1)\over\kappa}
\label{E:regconkappa}
\end{equation}
so, excluding the uninteresting possibility $r_H=0$ for
$d=1$, we see immediately that there can be no regular flow
with an acoustic horizon when $\kappa\leq 0$. For
$\kappa>0$, one can see that there are no values of $v_0$
for which (\ref{E:regconkappa}) is satisfied. Indeed,
setting $v=c$ and $r=r_H=c^2(d-1)/\kappa$ in
(\ref{E:diopoi}) gives the following equation for $v_0^2$:
\begin{equation}
W^{-1}\left(-{v_0^2\over c^2}\right)=-{1\over \rm
e}\left({\rm e}\,W^{-1}\left(-{\kappa r_0\over
c^2(d-1)}\right)\right)^{-2(d-1)}.
\label{E:ancorauna}
\end{equation}
Since, for $x<0$, it is $0>W^{-1}(x)>-1/{\rm e}$ (see figure
\ref{F:figlambert}), the right hand side of equation
(\ref{E:ancorauna}) turns out to be smaller than $-1/{\rm
e}$, while the left hand side is greater than $-1/{\rm e}$.
Therefore, satisfying equation (\ref{E:ancorauna}) is
impossible, {\em i.e.\/}, there are no real values of $v_0$
that satisfy it, and no regular flow exists in which $v=c$
at some point.

These conclusions are in agreement with equation
(\ref{E:d2}), which implies that $\d v/\d r$ cannot be
finite at $r_H$, because $\d^2\Phi/\d r^2\equiv 0$ in this
case. Thus, either $v(r)\neq c$ for all values of $r$, or
$\d v/\d r$ diverges at the horizon. It is not difficult to
see that the second possibility is the correct one, because
a horizon {\em always\/} forms in this type of flow. To this
end, let us set $v=c$ in (\ref{E:diopoi}), and look for a
solution $r_H$ (that we do {\em not\/} require to be
necessarily equal to the one following from the regularity
condition, equation (\ref{E:regconkappa}) --- in fact, we
already know that this would be impossible). We get
\begin{equation}
W^{-1}\left(-{\kappa r_H\over
c^2(d-1)}\right)=W^{-1}\left(-{\kappa r_0\over
c^2(d-1)}\right)
\left(-W^{-1}\left(-{v_0^2\over c^2}
\right)\right)^{1\over 2(d-1)}.
\end{equation}
The last factor on the right hand side of this equation is
always positive and smaller than one, therefore
\begin{equation}
W^{-1}\left(-{\kappa r_H\over c^2 (d-1)}\right)
\label{E:0}
\end{equation}
has the same sign of, and smaller absolute value than,
\begin{equation}
W^{-1}\left(-{\kappa r_0\over c^2 (d-1)}\right).
\label{E:H}
\end{equation}
For $\kappa<0$, (\ref{E:0}) is positive, so also (\ref{E:H})
is positive and corresponds to a positive value $r_H<r_0$.
For $\kappa>0$, (\ref{E:0}) is negative, so (\ref{E:H})
gives {\em two\/} positive solutions for $r_H$, one smaller
and the other greater than $r_0$. In both cases, the horizon
forms.

We now illustrate these features by showing some plots of
the solution of equation (\ref{E:d>1Kcos2}) with arbitrarily
chosen boundary conditions. Without loss of generality we
can rescale the unit of distance to set
\begin{equation}
\kappa = \left\{
\begin{array}{c}+ c^2/r_0,\\0,\\-c^2/r_0.\end{array}
\right.
\end{equation}
Let us treat these three cases separately.

\subsubsection{$\kappa>0$}

\begin{figure}[htbp]
\vbox{
\hfil
\scalebox{0.25}{{\includegraphics{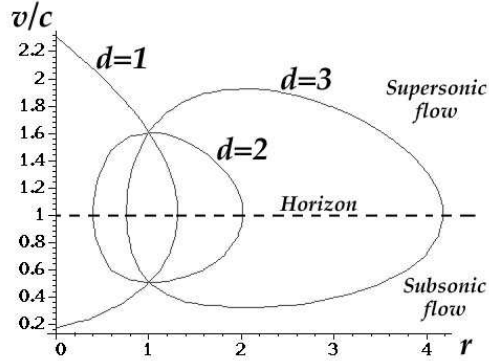}}}
\hfil
}
\bigskip
\caption{
Plot of the solutions of equation (\ref{E:d>1Kcos}) for
several values of $d$ and $\kappa>0$. We have first fixed
$\kappa=+c^2/r_0$, and then set $c=1$, $r_0=1$, and
$v_0=1/2$.
}
\label{F:d>1K}
\end{figure}

\begin{figure}[htbp]
\vbox{
\hfil
\scalebox{0.25}{{\includegraphics{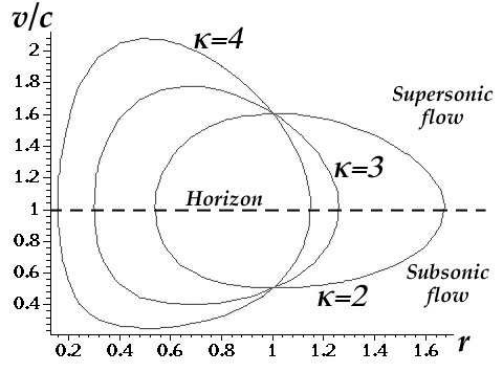}}}
\hfil
}
\bigskip
\caption{
Plot of the solutions of equation (\ref{E:d>1Kcos}) for
$\kappa=2,3,4$ with $d=3$, $c=1$, $r_0=1$, and $v_0=1/2$.
}
\label{F:d=3K>1}
\end{figure}

For $\kappa > 0$ figure \ref{F:d>1K} clearly shows that
there is no obstruction to reaching the acoustic horizon. In
addition, if we keep the distance scale fixed and instead
vary $\kappa$ we find the curves of figure \ref{F:d=3K>1}.


The four things to emphasize here are that:
\begin{enumerate}
\item
Velocities equal to the speed of sound are indeed attained;
\item
The gradient $dv/dr$ is indeed infinite at the acoustic
horizon;
\item
These particular solutions break down {\em at\/} the
acoustic horizon and cannot be extended {\em beyond\/} it;
\item
The particular solutions we have obtained all exhibit a
double-valued behaviour, there is a branch with subsonic
flow that speeds up and reaches $v=c$ at the acoustic
horizon; and there is a second supersonic branch, defined on
the same spatial region, that slows down and reaches $v=c$
at the acoustic horizon. Mathematically, this happens
because of the double-valuedness of the Lambert function of
negative argument, as already noted in section \ref{S:css}.
\end{enumerate}

\subsubsection{$\kappa=0$ (no body force)}

\begin{figure}[htbp]
\vbox{
\hfil
\scalebox{0.25}{{\includegraphics{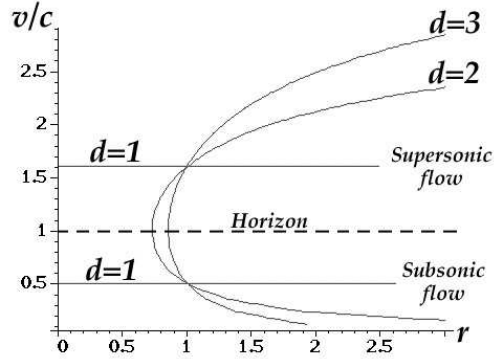}}}
\hfil
}
\bigskip
\caption{
Plot of the solution of equation (\ref{E:d>1Kcos}) for
$\kappa=0$ with the same initial values. ($c=1$, $r_0=1$,
and $v_0=1/2$.) Note the triviality of the $d=1$ solution,
which exhibits two branches, with subsonic and supersonic
speeds respectively.
}
\label{F:d>1}
\end{figure}

If there is no external body force, then $d=1$ is
uninteresting (the velocity is constant). If we now look at
$d=2$ and higher then equation (\ref{E:accsph}) again easily
gives us the acceleration of the fluid
\begin{equation}
a = v \; {\d v\over\d r} =
- v^2 \;
{
c^2 {(d-1)/r}
\over
c^2 - v^2 }\;,
\end{equation}
so
\begin{equation}
{\d v\over\d r} =
- v \;
{
c^2 {(d-1)/r}
\over
c^2 - v^2
}.
\end{equation}
Explicit integration leads us to the solution
\begin{equation}
r=r_0\; \left(\frac{v}{v_0}\right)^{\frac{1}{1-d}}
\exp\left(\frac{v^2-v^2_0}{2(d-1)c^2}\right),
\end{equation}
which is equivalent to equation (\ref{E:diopoi}) in the
limit $\kappa=0$. This can be easily plotted for different
values of the dimension $d$ as shown in figure \ref{F:d>1}.

\subsubsection{$\kappa<0$}

\begin{figure}[htbp]
\vbox{
\hfil
\scalebox{0.25}{{\includegraphics{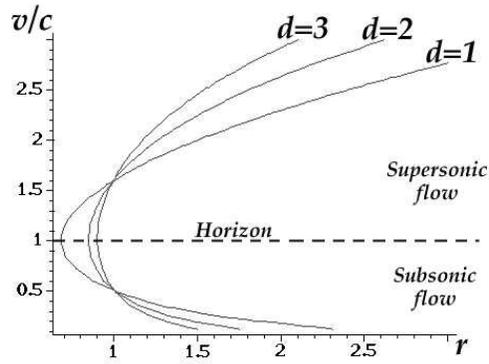}}}
\hfil
}
\bigskip
\caption{
Plot of the solution of equation (\ref{E:d>1Kcos}) for
$d=1,2,3$, and $\kappa=-c^2/r_0$, where we again set $c=1$,
$r_0=1$, and $v_0=1/2$.
}
\label{F:k=-1}
\end{figure}

For $\kappa<0$ the solutions are plotted in figure \ref{F:k=-1}.

\begin{figure}[htbp]
\vbox{
\hfil
\scalebox{0.25}{{\includegraphics{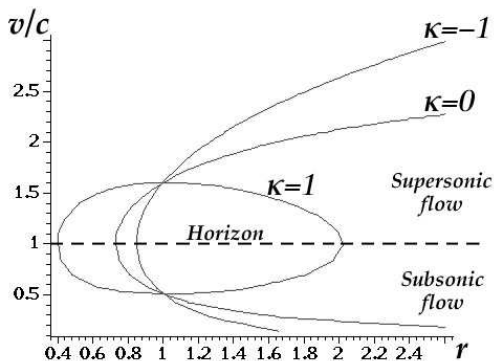}}}
\hfil
}
\bigskip
\caption{
Plot of the solutions of equation (\ref{E:d>1Kcos}) for
$\kappa=\pm1,0$ and $d=2$, with $c=1$, $r_0=1$, and
$v_0=1/2$.
}
\label{F:d2k101}
\end{figure}

Finally it is interesting to compare the behaviour of the
solutions for the different signs of the body force as shown
in figure \ref{F:d2k101}.

In all three cases ($\kappa>0$, $\kappa=0$, $\kappa<0$) we
see that the acoustic horizon does in fact form as
predicted, and that the surface gravity and acceleration are
indeed infinite at the acoustic horizon. Naturally, this
should be viewed as evidence that some of the technical
assumptions usually made are no longer valid as the horizon
is approached. In particular in the next section (section
\ref{S:viscosity}) we shall discuss the role of viscosity as
a regulator for keeping the surface gravity finite.

\subsection{Schwarzschild geometry}

So far, the discussion in this paper has concerned the
attainability of acoustic horizons in general, without
focusing on any particular acoustic geometry. A more
specific, and rather attractive possibility is to attempt to
build a flow with an acoustic metric that is as close as
possible to one of the standard black hole metrics of
general relativity. Remarkably, this can be done (up to a
conformal factor) for the Schwarzschild geometry. To be more
specific: for a fluid with constant speed of sound, one can
find a stationary, spherically symmetric flow in three
spatial dimensions, whose acoustic metric is conformal to
the {\Painleve}--Gullstrand form of the Schwarzschild
geometry~\cite{Visser98}. This possibility has stimulated
considerable work concerning the physical realization of an
experimental setup that could actually produce such a flow
(or, more precisely, a two-dimensional version of
it~\cite{Volovik:1999fc}). These particular fluid
configurations exhibit a different type of fine-tuning
problem than the one we discussed previously. In order to
reproduce the {\Painleve}--Gullstrand line element, the
speed of the fluid must have the profile $v=\sqrt{2M/r}$,
with $M$ a positive constant. Then, $r_0$ and $v_0$ must
satisfy the relation $r_0v_0^2=2M$, and equation
(\ref{E:d>1Kcos}) allows us to find the external potential
needed in order to sustain such a flow in $d$ space
dimensions:
\begin{equation}
\Phi(r)=\Phi(r_0) + {M\over r_0} +
c^2\left(d-{3\over 2}\right)\ln\left({r\over r_0}
\right)-{M\over r}.
\label{E:pgpot}
\end{equation}
Therefore, the potential must be carefully chosen, which
will not be easy to do in a laboratory. If one does manage
to construct such a potential $\Phi(r)$ it will
automatically fulfill the fine-tuning condition
(\ref{E:fine-tuning}) at the acoustic horizon, $r_H=2M/c^2$.
This is only to be expected, because $\d v/\d
r=-\sqrt{M/2r^3}$ blows up only as $r\to 0$. Also, since we
know the surface gravity of a Schwarzschild black hole is
finite, any fluid flow that reproduces the Schwarzschild
geometry must by definition satisfy the fine tuning
condition for a finite surface gravity.

Looking at the issue from the point of view discussed at the
end of section \ref{S:css}, one expects that, given the
potential (\ref{E:pgpot}), the value $v_0=\sqrt{2M/r_0}$ is
the solution of equations (\ref{E:rH}) and (\ref{E:regcon}),
while $v(r)=\sqrt{2M/r}$ is the corresponding eigenfunction
that is selected by the requirement of having a regular
flow. This is indeed the case: Equation (\ref{E:rH}) now
gives $r_H=2M/c^2$ which, substituted into (\ref{E:regcon}),
leads to the following equation for $v_0$:
\begin{equation}
{c^2\over 2}\,\ln\left({2M\over r_0v_0^2}\right)+{v_0^2\over
2}={M\over r_0}.
\label{E:v0pg}
\end{equation}
It is trivial to check that $v_0=\sqrt{2M/r_0}$ is, in fact,
a solution of (\ref{E:v0pg}).

\begin{figure}[htbp]
\vbox{
\hfil
\scalebox{0.25}{{\includegraphics{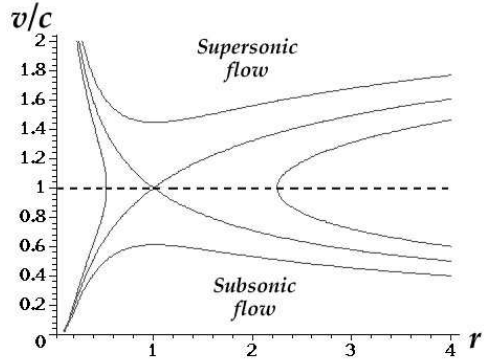}}}
\hfil
}
\bigskip
\caption{
Plot of the solutions of equation (\ref{E:d>1Kcos}) for the
potential (\ref{E:pgpot}), with $M=1/2$, $d=3$, $c=1$, and
$r_0=4$. The corresponding boundary conditions are
$v_0=0.4$, $v_0=0.5$, and $v_0=0.6$.
}
\label{F:pg}
\end{figure}

Considering the same potential (\ref{E:pgpot}), but values
of $v_0$ different from $\sqrt{2M/r_0}$, corresponds to
flows either with no horizon, or in which $(\d v/\d r)_H$
diverges. This is evident in figure \ref{F:pg}, which
confirms the ``eigenvalue character'' of the problem of
finding a regular flow. Notice that there are two solutions
that are regular at the horizon, with opposite values of
$(\d v/\d r)_H$, in full agreement with the fact that
equation (\ref{E:dvdrh}) only determines the {\em square\/}
of $(\d v/\d r)_H$.

Additionally, note that what we have done above has been to
ask how to mimic a slice of the $(3+1)$-dimensional
Schwarzschild geometry with a $(d+1)$-dimensional fluid
flow. We could ask what happens in different spacetime
dimensions: for the $(D+1)$-dimensional generalization of
the Schwarzschild geometry the fluid flow generalizes to
$v=\sqrt{2M/r^{D-2}}$, and the potential gradient required
to produce this flow is
\begin{equation}
\Phi(r)=
\Phi(r_0) + {M\over r_0^{D-2}} +
c^2\left(d-{D\over 2}\right)\ln\left({r\over r_0}\right)-
{M\over r^{D-2}}.
\end{equation}
Again a very specific external body force is needed to set
up the very specific fluid flow corresponding to a
higher-dimensional Schwarzschild geometry.

\subsection{Reissner--{\Nordstrom} geometry}

We mention in passing that generalizing this discussion to
the $(3+1)$-dimensional Reissner--{\Nordstrom} geometry is
straightforward. This geometry is described in
{\Painleve}--Gullstrand form by the fluid flow
\begin{equation}
v(r) = \sqrt{ {2M \over r} - {e^2\over r^2} }.
\end{equation}
The external potential required to set up this fluid flow is
then
\begin{equation}
\Phi(r)=
c^2\left(d-{3\over 2}\right)\ln\left({r\over r_0}\right)-
{M\over r} + {e^2\over 2r^2}+{c^2\over 2}\ln
\left( 1 - {e^2\over 2Mr}\right)+\mbox{const}.
\end{equation}
%

\subsection{The canonical acoustic black hole}

To wrap up our section on specific examples, we add a few
words about the ``canonical'' acoustic black hole discussed
in~\cite{Visser98,Visser99}. In that model the fluid is
assumed to have a constant density throughout space, and the
continuity equation is then used to deduce the velocity
profile
\begin{equation}
v(r) \propto 1/r^2.
\end{equation}
Note that ``constant density'' is actually a much weaker
statement than incompressibility, and the word
incompressible should be excised from all of section 8
of~\cite{Visser98} and replaced by this phrase. Now
in~\cite{Visser98,Visser99} the velocity profile was
determined purely on these kinematic grounds, and no attempt
was made to put this background fluid flow back into the
Euler equations to determine the external body force
required to set up the flow. (In that paper, almost all the
attention was focussed on the fluctuations rather than the
background flow.)

Determining the potential is an easy application of the
general analysis of this article [see
equation~(\ref{E:glu})]. We calculate
\begin{equation}
\Phi(r) \propto  - 1/r^4+\mbox{const}.
\end{equation}
With hindsight this can be seen to be nothing more than a
special case of Bernoulli's theorem for a constant-density
flow
\begin{equation}
\Phi(r) =  - {1\over2} \; v^2+\hbox{const}.
\end{equation}

The single over-riding message coming from all these specific examples
is the generic dichotomy between a formally infinite surface gravity
and needing a highly specific boundary condition to be satisfied. In
the following section we shall regulate the generically infinite
surface gravity by using a less idealized model for the fluid.

\section{Viscosity}
\setcounter{equation}{0}
\label{S:viscosity}

A viscous flow is governed by the Navier-Stokes equation
(\ref{E:euler}). In general, there will be two contributions to
viscosity, associated with the coefficients $\eta$ and $\zeta$. Since
our treatment in the present section does not pretend to be realistic,
but we simply wish to point out how viscosity acts as a regulator for
the surface gravity, we shall set the bulk viscosity coefficient
$\zeta$ to zero, in order to have a model with as few free
parameters as possible. (This is sometimes called the ``Stokes
assumption'' \cite{kundu}.) For a spherically symmetric inflow one has
\begin{equation}
\vec{n}\cdot\left(\nabla^2\v+{1\over
3}\,\vec{\nabla}\vec{\nabla}\cdot\v\right)={4\over
3}\,\left({\d^2 v\over\d r^2}+{(d-1)\over r}{\d v\over\d
r}-{(d-1)v\over r^2}\right),
\end{equation}
so the Navier-Stokes equation (\ref{E:euler}) becomes, in
the stationary case,
\begin{eqnarray}
a = v \; {\d v\over\d r} &=&
- {v^2 \over c^2 - v^2}\;
\left[{c^2(d-1)\over r}-{\d\Phi\over\d r}\right.\nonumber\\
&+& \left. {4\nu\over 3}\left({\d^2 v\over\d
r^2}+{(d-1)\over r}{\d v\over\d r}-{(d-1)v\over
r^2}\right)\right],
\label{E:nss}
\end{eqnarray}
where we have introduced the coefficient of kinematic viscosity
$\nu=\eta/\rho$. Hereafter we assume, for simplicity, that $\nu$ is a
constant. (However, any hypothesis about $\nu$ must ultimately by
justified by a kinetic model for the fluid, and it is worth noticing
that there are plausible distribution functions that lead to a
velocity-dependent $\nu$; see, \eg, \cite{Narayan}. Obviously, such a
dependence can have important repercussions on the conclusions of the
present section.) The acceleration is then infinite at the horizon
unless
\begin{equation}
\left(1-{4\nu\over 3c r_H}\right){c^2(d-1)\over r_H} -
\left.{\d\Phi\over\d r}\right|_H +
{4\nu\over 3}\left(\left.{\d^2 v\over dr^2}\right|_H
+{(d-1)\over r_H}\left.{\d v\over\d r}\right|_H\right)= 0.
\end{equation}
Since this involves higher-order derivatives at the horizon,
it can no longer be regarded as a fine-tuning constraint, or
as an equation for $r_H$, but merely as a statement about
the shape of the velocity profile near the horizon. Indeed
the general solution to this equation when $c$ is constant
is
\begin{eqnarray}
v(r) &=& c + {a_H\over c} (r-r_H) + {(d-1)\over 2cr_H}
\left({c^2\over r_H}-a_H\right)(r-r_H)^2\nonumber\\
&+&{3\over 8\nu}\left(\left.{\d\Phi\over\d r}\right|_H
-{c^2(d-1)\over r_H}\right)
(r-r_H)^2 + O\left((r-r_H)^3\right),
\end{eqnarray}
with $a_H$ arbitrary, and finite, at the horizon.  We
can rearrange (\ref{E:nss}) to get a differential equation
for $v(r)$:
\begin{equation}
{\d^2 v\over\d r^2} +\left({(d-1)\over r}+{3\over
4\nu}{c^2-v^2\over v}\right){\d v\over\d r}-{(d-1)v\over
r^2}= {3\over 4\nu}
\left({\d\Phi\over\d r} - {c^2(d-1)\over r}\right).
\end{equation}
Unfortunately, integrating this equation is completely impractical in
general and we must resort to the analysis of special cases.

\subsection{$d=1$, constant body force}

\begin{figure}[htbp]
\vbox{
\hfil
\scalebox{0.30}{{\includegraphics{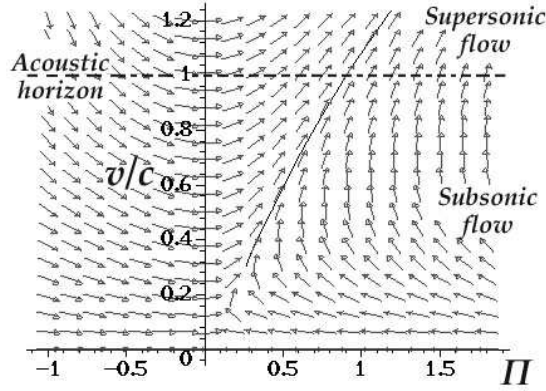}}}
\hfil
}
\bigskip
\caption{
Plot of the phase space for $d=1$, $\kappa=\mbox{constant} >
0$, and nonzero viscosity. The transverse line identifies a
separatrix in the integral curves. Note that the integral
curves can intersect the acoustic horizon at arbitrary
values of the surface gravity.
}
\label{F:phase}
\end{figure}

Even the case of constant body force is intractable unless
$d=1$, in which case we get (following the steps above)
\begin{equation}
{\d v\over\d r} =
- {v\over c^2 - v^2}  \left(
-\kappa + {4\nu\over 3} {\d^2 v\over\d r^2} \right).
\label{E:difficeq}
\end{equation}
This single second-order differential equation can be turned
into an {\em autonomous\/} system of first-order equations
\begin{equation}
\left \{ \begin{array}{lll}
\displaystyle{\frac{\d v}{\d r}}&=&\Pi, \\
&&\\
\displaystyle{\frac{\d \Pi}{\d r}}
&=&\displaystyle{
\frac{3\kappa}{4\nu}+\frac{3v}{4\nu}\left(1-
\frac{c^2}{v^2}\right)\Pi}.
\end{array}
\right.
\end{equation}
We can plot the flow of this autonomous system in the usual
way and it clearly shows that it is possible to cross the
acoustic horizon $v=c$ at arbitrary accelerations $a_H$ and
arbitrary surface gravity $g_H$ (see figure \ref{F:phase}).

\subsection{$d=1$, zero body force}

Integrating equation (\ref{E:difficeq}) once (this is easy provided
$c$ is a constant), we get
\begin{equation}
\left.{\d v\over\d r}\right|_r =
\left.{\d v\over\d r}\right|_{r_0}
-
{3\over8\nu} \;
\left[
c^2 \;
\ln\left( \frac{v^2}{v^2_0}\right)-v^2+v^2_0
\right]+{3\over 4\nu}\left(\Phi(r)-\Phi(r_0)\right),
\end{equation}
where $(r_0,v_0)$ again denotes an arbitrary pair of initial
values. If $d=1$ and $\kappa =0$ this equation for $\d v(r)/\d r$
reduces to
\begin{equation}
\left.{\d v\over\d r}\right|_r =
\left.{\d v\over\d r}\right|_{r_0}
-
{3\over 8\nu} \;
\left[
c^2 \;
\ln\left( \frac{v^2}{v^2_0}\right)-v^2+v^2_0
\right].
\end{equation}
In this particular case the analysis is sufficiently simple
that we can say something about the acceleration at the
horizon, namely
\begin{equation}
\left.{\d v\over\d r}\right|_H =
\left.{\d v\over\d r}\right|_{r_0}
-
{3\over8\nu} \;
\left[
c^2 \;
\ln\left( \frac{c^2}{v^2_0}\right)-c^2+v^2_0
\right].
\end{equation}
That is
\begin{equation}
g_H = a_H = c \left.{\d v\over\d r}\right|_{r_0}
-
{3c\over8\nu} \;
\left[
c^2 \;
\ln\left( \frac{c^2}{v^2_0}\right)-c^2+v^2_0
\right],
\end{equation}
which is an explicit analytic verification that viscosity
regularizes the surface gravity of the acoustic horizon.

\begin{figure}[htbp]
\vbox{
\hfil
\scalebox{0.30}{{\includegraphics{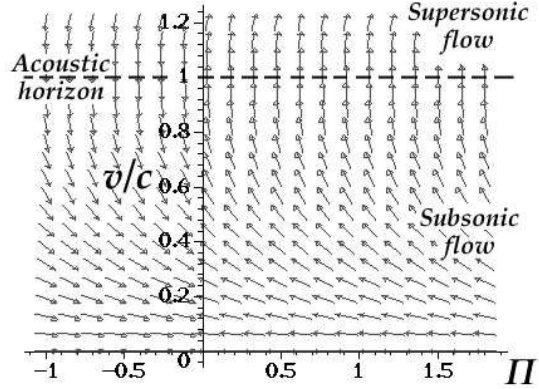}}}
\hfil
}
\bigskip
\caption{
Plot of the phase space for $d=1$, $\kappa=0$, and nonzero
viscosity. Again note that the integral curves can intersect
the acoustic horizon at arbitrary values of the surface
gravity.
}
\label{F:phasek0}
\end{figure}

We can plot the flow in the usual way and it again clearly
shows that it is possible to cross the acoustic horizon
$v=c$ at arbitrary accelerations $a_H$ (see figure
\ref{F:phasek0}).

\subsection{$d>1$, constant body force}

The relevant equation is
\begin{equation}
{\d^2 v\over\d r^2} +\left({(d-1)\over r}+{3\over
4\nu}{c^2-v^2\over v}\right){\d v\over\d r}-{(d-1)v\over
r^2}= {3\over 4\nu}
\left(\kappa - {c^2(d-1)\over r}\right),
\end{equation}
which can be recast as
\begin{equation}
\left \{ \begin{array} {lll}
\displaystyle{{\d v \over \d r}} &=&\Pi \\
&&\\
\displaystyle{\frac{\d \Pi}{\d r}}
&=& {\displaystyle {3\over
4\nu}\left(\kappa-\frac{c^2(d-1)}{r}\right)+ {(d-1)v\over
r^2} +\left(\frac{3v}{4\nu}\left(1-\frac{c^2}{v^2}\right)-
{(d-1)\over r}\right)\Pi}.
\end{array}
\right.
\label{E:non-autonomous}
\end{equation}
This is no longer an {\em autonomous} system of differential
equations, (there is now an explicit $r$ dependence in these
equations) so a flow diagram is meaningless. Nonetheless the
system can be treated numerically and curves plotted as a
function of initial conditions. As an example we plot some
curves in the phase space for $d=2$, and verify that at
least some of these curves imply formation of an acoustic
horizon.

\begin{figure}[htbp]
\vbox{
\hfil
\scalebox{0.30}{{\includegraphics{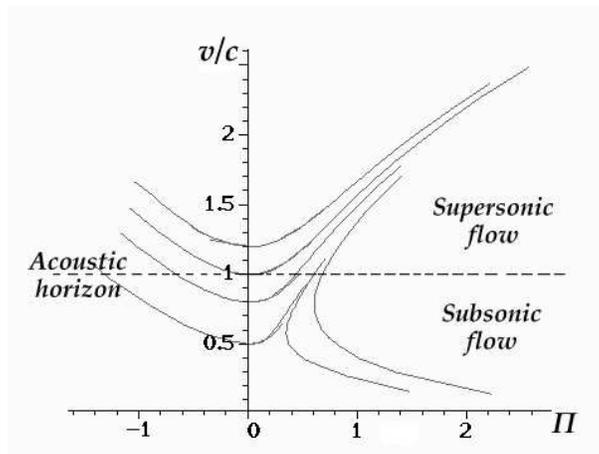}}}
\hfil
}
\bigskip
\caption{
Plot of some solutions to the non-autonomous system
(\ref{E:non-autonomous}) for various initial conditions for
$d=2$, $\kappa=\mbox{constant}$, and nonzero viscosity. Note
that at least some of these curves intersect the acoustic
horizon, and do so at various but finite values of the
surface gravity.
}
\label{phased>1}
\end{figure}

As a final remark we think it is useful to briefly discuss
the effect of viscosity with regard to Hawking radiation. It
has been shown that the addition of viscosity to the fluid
dynamical equations is equivalent to the introduction of an
explicit violation of ``acoustic Lorentz invariance'' at
short scales~\cite{Visser98,Visser97}. Thus one may wonder
if such an explicit breakdown would not lead to a
suppression of the Hawking flux as well. Indeed the
violation of Lorentz invariance is important for wavelengths
of order $\lambda_0 =
\nu/c$~\cite{Visser98,Visser97}, introducing in this way a
sort of cutoff on short wavelengths which can dramatically
affect the Hawking
flux~\cite{Jacobson91,Jacobson93,Unruh94,Mini-survey}.

Thus there is naively a risk that using viscosity to remove
the unphysical divergences at acoustic horizons would also
``kill'' the phenomenon one is seeking. This problem has
been extensively discussed in the literature
(see~\eg~\cite{Mini-survey}) and it has (quite remarkably)
been demonstrated that such a violation of Lorentz
invariance is not only harmless but even natural and useful.
In particular, viscosity can be shown to
induce~\cite{Visser98,Visser97} the same type of
modifications of the phonon dispersion relation which are
actually required for circumventing the above cited
problem~\cite{Jacobson91,Jacobson93,Unruh94,Mini-survey}. So
the emergence of viscosity appears to be indeed a crucial
factor, both for allowing the formation of acoustic horizons
and, at the same time, for implementing that mechanism of
``mode regeneration'' which permits Hawking radiation in
presence of short distance cutoff.

\section{Conclusions: Danger+opportunity.}
\setcounter{equation}{0}

Let us summarize the results that we have obtained for a fluid
subjected to a given external potential: In the viscosity-free,
stationary case, we have seen that if the flow possesses an acoustic
horizon, the gradients of physical quantities, as well as the surface
gravity and the corresponding Hawking flux, generically exhibit formal
divergences. There are two ways in which a real fluid can circumvent
this physically unpalatable result. For a broad class of potentials,
there is one particular flow which is regular everywhere, even at the
horizon. In this case, it is obvious that the fluid itself will
``choose'' such a configuration. Mathematically, imposing the
regularity condition at the horizon amounts to formulating an
eigenvalue problem. However, there are physically interesting
potentials --- such as a linear one --- for which this is
impossible. We view this result as both a danger and an opportunity. A
danger because infinite accelerations are clearly unphysical and
indicate that the idealization of considering a {\em irrotational
barotropic inviscid perfect fluid\/} (and this idealization underlies
the standard derivations of the notion of acoustic
metric~\cite{Unruh81,Visser93,Visser98,Visser99}), is sure to break
down in the neighborhood of any putative acoustic horizon. Indeed, the
divergences will be avoided in real life simply because one or more of
the simplifying hypothesis become invalid as $v\to c$.

On the other hand, this may be viewed as an opportunity: Once we
regulate the infinite surface gravity, by adding for instance a finite
viscosity, we find that the surface gravity becomes an extra free
parameter, divorced from naive estimates based on the geometry of the
fluid flow. The common naive estimates of the surface gravity take the
form~\cite{Unruh81,Visser99}
\begin{equation}
g_H \approx {c^2\over R},
\end{equation}
where $R$ is a typical length scale associated with the flow (a nozzle
radius, or the radius of curvature of the horizon). The analysis of
this note suggests that this estimate may in general be misleading
because it does not take into account information regarding the
dynamics of the flow. Because of this the surface gravity could be
considerably larger than previously expected.

There is a potential source of confusion which we should clarify
before wrapping up: in general relativity the physical acceleration of
a {\em stationary\/} observer hovering just outside the black hole
horizon diverges, but when an appropriate red-shift factor is applied
and the properly defined surface gravity is calculated that surface
gravity proves to be finite. On the contrary, in the acoustic black
holes it is the Newtonian acceleration of the {\em infalling\/}
observers that (in the absence of fine-tuning) diverges at the
horizon, leading to an infinite surface gravity. Why the difference?
It is here that the actual dynamical equations governing the
background geometry come into play. The physics that is identical
between gravitational black holes and acoustic black holes is the
kinematical physics of fields propagating in the respective Lorentzian
spacetime metrics. The physics which is different is that which
depends on the dynamical equations of motion of the background
geometry. For gravity, the latter is governed by the Einstein
equations while for acoustic black holes it is governed by the
hydrodynamic equations --- these equations are sufficiently different
that the geometries of the two Lorentzian metrics can be quite
different, even though qualitative features such as the existence of
event horizons may be quite similar.

Our general discussion, plus the specific example utilizing viscosity,
makes it clear that it is the specific technical restrictions placed
on the hydrodynamic equations that lead to the formally infinite
surface gravity --- and so one might wonder how much of the current
analysis to trust. For example, in real superfluids the existence of
{\em roton\/} excitations leads to a breakdown of irrotational flow
before the acoustic horizon is reached~\cite{Mini-survey}. Adding
vorticity is certainly technically complicated (see for instance the
recent book by Ostashev~\cite{Ostashev}), but this may merely be a
technical complication, not a fundamental barrier to progress. For
technical discussions regarding the possibility (probability) of
actually building acoustic black holes
see~\cite{Volovik:1997xi,Kopnin:1998jy,Volovik:1998de,Volovik:1997pf,%
Jacobson:1998ms,Jacobson:1998he,Volovik:1999zs,Volovik:1999fc,%
Volovik:1999cn,Volovik:2000,Garay,Mini-survey}. Note that the Garay
{\etal} implementation of acoustic black holes \cite{Garay} is built
on a different physical background; they use Bose--Einstein condensate
governed by the Gross--Pitaevski equation rather than a barotropic
fluid governed by the Euler-continuity equations. Therefore the perils
and opportunities delineated in this article do not necessarily apply
to their particular situation. A similar remark applies to Volovik's
implementation based on two-fluid models of superfluidity (for example
${}^3$He-A), where the horizon is defined using the speed of the
quasi-particles, rather than by the speed of sound {\em per se}.  Of
these two speeds, the former is much smaller than the latter, so the
surface gravity at such horizons is always
finite~\cite{Volovik:2000}. In short, while specific physical
implementations of the acoustic geometry idea all have their
characteristic peculiarities and potential pitfalls, overall the
experimental prospects continue to look extremely promising.

\section*{Acknowledgements}

We are grateful to Grisha Volovik for a remark that stimulated an
improvement in the presentation. MV would like to thank Rob Myers and
Ted Jacobson for some penetrating questions and useful discussion. SL
and SS are grateful to John Miller and Ewa Szuszkiewicz for calling
their attention on reference~\cite{Narayan}. SL acknowledges
hospitality and financial support from the Washington University in
Saint Louis, where part of this work was performed. The work of MV was
supported by the US Department of Energy. MV also wishes to
acknowledge hospitality and financial support from SISSA, Trieste.



\begin{thebibliography}{99}
\bibitem{Unruh81}
W.G. Unruh, ``Experimental black hole evaporation?'' Phys.
Rev. Lett.\, {\bf 46}, 1351--1353 (1981).
\bibitem{Jacobson91}
T.A. Jacobson, ``Black hole evaporation and ultrashort
distances'', Phys. Rev. D {\bf 44}, 1731--1739 (1991).
\bibitem{Jacobson93}
T.A. Jacobson, ``Black hole radiation in the presence of a
short distance cutoff'', Phys. Rev. D {\bf 48}, 728--741
(1993) [hep-th/9303103].
\bibitem{Visser93}
M. Visser, ``Acoustic propagation in fluids: an unexpected
example of Lorentzian geometry'', gr-qc/9311028.
\bibitem{Unruh94}
W.G. Unruh, ``Sonic analogue of black holes and the
effects of high frequencies on black hole evaporation'',
Phys. Rev. D {\bf 51}, 2827--2838 (1995)
[gr-qc/9409008].
\bibitem{Visser98}
M. Visser, ``Acoustic black holes: Horizons, ergospheres,
and Hawking radiation'', Class. Quantum Grav. {\bf 15},
1767--1791 (1998) [gr-qc/9712010].
\bibitem{Visser97}
M. Visser, ``Hawking radiation without black hole entropy'',
Phys. Rev. Lett. {\bf 80}, 3436--3439 (1998) [gr-qc/9712016].
\bibitem{Visser99}
M. Visser, ``Acoustic black holes'',
Lecture delivered at the Advanced School on Cosmology
and Particle Physics, Peniscola, Spain, June 1998;
gr-qc/9901047.
\bibitem{Volovik:1997xi}
G.E. Volovik, ``Simulation of quantum field theory and
gravity in superfluid ${}^3$He'', Low Temp. Phys. (Kharkov) {\bf
24}, 127--129 (1998) [cond-mat/9706172].
\bibitem{Kopnin:1998jy}
N.B. Kopnin and G.E. Volovik, ``Critical velocity and event
horizon in pair-correlated systems with ``relativistic''
fermionic quasiparticles'', Pisma Zh. Eksp. Teor. Fiz.\ {\bf
67}, 124--129 (1998) [cond-mat/9712187].
\bibitem{Volovik:1998de}
G.E. Volovik, ``Gravity of monopole and string and
gravitational constant in ${}^3$He-A'', Pisma Zh. Eksp.
Teor.\ Fiz. {\bf 67}, 666--671 (1998); JETP Lett. {\bf 67},
698--704 (1998) [cond-mat/9804078].
\bibitem{Volovik:1997pf}
G.E. Volovik, ``Induced gravity in superfluid ${}^3$He'',
J. Low Temp. Phys.  {\bf 113}, 667--680 (1997) [cond-mat/9806010].
\bibitem{Jacobson:1998ms}
T.A. Jacobson and G.E. Volovik, ``Event horizons and
ergoregions in ${}^3$He'', Phys. Rev. D {\bf 58}, 064021 (1998).
\bibitem{Jacobson:1998he}
T.A. Jacobson and G.E. Volovik, ``Effective spacetime and
Hawking radiation from moving domain wall in thin film of
${}^3$He-A'', Pisma Zh. Eksp. Teor. Fiz. {\bf 68}, 833--838
(1998); JETP Lett. {\bf 68}, 874--880 (1998) [gr-qc/9811014].
\bibitem{Volovik:1999zs}
G.E. Volovik, ``Field theory in superfluid ${}^3$He: What
are the lessons for particle physics, gravity, and high
temperature superconductivity?'' Proc. Nat. Acad. Sci.\ {\bf
96}, 6042--6047 (1999) [cond-mat/9812381].
\bibitem{Volovik:1999fc}
G.E. Volovik, ``Simulation of \Painleve--Gullstrand black
hole in thin ${}^3$He-A film'', Pisma Zh. Eksp. Teor. Fiz.\
{\bf 69}, 662--668 (1999); JETP Lett. {\bf 69} 705--713
(1999) [gr-qc/9901077].
\bibitem{Volovik:1999cn}
G.E. Volovik, ``${}^3$He and universe parallelism'', in 
{\em Topological defects and the Non-Equilibrium Dynamics of 
Symmetry Breaking Phase Transitions"}, 
Eds. Y.M. Bunkov and H. Godfrin, Kluwer Academic Publishers, 2000, 
pp. 353 - 387;
[cond-mat/9902171].
\bibitem{Volovik:2000} G.E. Volovik, ``Links between
gravity and dynamics of quantum liquids'', gr-qc/0004049.
\bibitem{Garay}
L.J. Garay, J.R. Anglin, J.I. Chirac and P. Zoller, ``Black
holes in Bose--Einstein condensates'', gr-qc/0002015.
\bibitem{Mini-survey}
T.A. Jacobson, ``Trans--Planckian redshifts and substance of
the spacetime river'', hep-th/0001085.
\bibitem{Nielsen78}
H.B. Nielsen and M. Ninomiya, ``Beta function in a
non-covariant Yang-Mills theory'', Nucl. Phys. {\bf B141},
153--177 (1978).
\bibitem{Nielsen83a}
H.B. Nielsen and I. Picek, ``Lorentz non-invariance'', Nucl.
Phys. {\bf B211}, 269--296 (1983).
\bibitem{Nielsen83b}
S. Chadha and H.B. Nielsen, ``Lorentz invariance as a low
energy phenomenon'', Nucl. Phys. {\bf B217}, 125--144
(1983).
\bibitem{ll} L.D. Landau and E.M. Lifshitz, {\em Fluid
Mechanics\/} (Pergamon, Oxford, 1959).
\bibitem{kundu}
P.K. Kundu, {\em Fluid Mechanics\/} (Academic, 1990).
\bibitem{Lambert}
R.M. Corless, G.H. Gonnet, D.E.G. Hare, D.J. Jeffrey and
D.E. Knuth, ``On the Lambert $W$ function'', Adv. Comp.
Math. {\bf 5}, 329--359 (1996).
\bibitem{Narayan}
R. Narayan, ``A flux-limited model of particle
diffusion and viscosity'', Ap. J. {\bf 394}, 261--267
(1992).
\bibitem{Ostashev}
V.E. Ostashev, {\em Acoustics in Moving Inhomogeneous
Media\/} (E \& FN Spon, Thompson Professional, London,
1997).
\end{thebibliography}
\end{document}